\documentclass[journal=jctcce,manuscript=article]{achemso}

\usepackage[utf8]{inputenc}
\usepackage{amsmath, amssymb}
\usepackage[version=3]{mhchem} 
\usepackage{graphicx}

\usepackage{amsmath}

\usepackage{hyperref}

\usepackage{color}
\usepackage{subcaption}
\usepackage{comment}

\newcommand{\red}[1]{{\color{red} #1}}

\newcommand{\sibr}{SiBr$_4$}

\author{Zden\v{e}k Ma\v{s}\'{i}n, Jakub Benda, Martin Crh\'{a}n}
\affiliation[charles]
{Institute of Theoretical Physics, Faculty of Mathematics and Physics, Charles University, V Hole\v{s}ovi\v{c}k\'{a}ch 2, 180 00 Prague 8, Czech Republic}

\author{Gregory S. J. Armstrong, Anna Nelson, Sebastian Mohr}
\affiliation[quantemol]
{Quantemol Ltd., 320 City Road, Angel, London EC1V 2NZ, United Kingdom}

\author{Jonathan Tennyson}
\affiliation{Department of Physics and Astronomy, University College London, Gower Street, WC1E 6BT London, United Kingdom}
\alsoaffiliation[quantemol]
{Quantemol Ltd., 320 City Road, Angel, London EC1V 2NZ, United Kingdom}
\email{j.tennyson@ucl.ac.uk; zdenek.masin@matfyz.cuni.cz}
\title[ECPs]{Effective Core Potentials for calculations of continuum spectra of molecules using the molecular R-matrix method}

\date{\today}
\abbreviations{IR,NMR,UV}

\begin{document}


\begin{abstract}
  Implementation of Effective Core Potentials (ECPs) into the molecular scattering suite UKRmol+ is presented together with a set of calculations for a range of targets relevant for plasma modeling. Continuum description in scattering and photoionization calculations for large targets or high-energy electrons often requires the use of numerical continuum functions and the associated molecular integrals. We derive expressions for ECP integrals over B-spline type orbitals using their momentum-space representation and describe their implementation. Sample calculations are presented for electron collision from bromine molecule (Br$_2$), silicon tetrabromide (SiBr$_4$) and tungsten hydride (WH) as well as photoionisation of methyl iodide (CH$_3$I).
\end{abstract}

\section{Introduction} 

Electron dynamics in atomic and molecular systems is challenging because of the correlated motion of electrons but also due to the possible presence of relativistic effects in systems containing heavy atoms. In addition a practical complication arises in calculation of the related molecular two-electron integrals, which generally scales with the fifth power of the number of the molecular orbitals.

One way of alleviating these challenges is to take advantage of the orbital occupation model to reduce the number of electrons explicitly included in the molecular Hamiltonian replacing them with effective potentials, representing an average effect of those electrons on the remaining ones~\cite{reiher2014a,dolg2012}. Naturally, the core and inner-valence electrons can be satisfactorily represented by the mean field when dynamics occurring in the valence space is considered. Methods of this kind have been developed for bound state quantum chemistry calculations starting in 1960s. Since then they have become ubiquitous and standard part of bound state calculations. Two major strands of effective potentials have been developed: pseudopotentials and model potentials. The latter aims to preserve the correct nodal structure of the valence orbitals while the former leads to so called pseudovalence orbitals whose representation may benefit from using specialized Gaussian-type atomic bases~\cite{bettega1996a}. For that reason the choice of the pseudopotential often comes with a recommended optimal Gaussian-type valence basis set. The pseudopotentials, more commonly called Effective Core Potentials (ECPs), have become the dominant effective representation of the core. Since the late 1970s ECPs have been extended to incorporate scalar and spin-orbit relativistic effects of the core on the valence shells. The history and properties of the various forms of the effective and model potentials used to date have been described in comprehensive reviews~\cite{dolg2012,krauss1984}.

The use of ECPs in calculations of continuum spectra of molecules is much less common. To the best of our knowledge the only working implementation of those is in the Schwinger multi-channel codes of the Brazilian group~\cite{costa2015}. While ab initio multi-electron approaches for continuum states are able to represent electron correlation at a high level, their extension to inclusion of relativistic effects remains a challenge. Relativistic effects have been included in all-electron atomic scattering and photoionization calculations starting with the R-matrix implementation in the early 1980s~\cite{scott1980}. Fully relativistic atomic codes based on the Dirac equation also exist, see e.g.~\cite{Bostock_2011,zats2016}. More recently relativistic effects have been included in time-dependent simulations of atoms in external fields~\cite{brown2019} but their extension to molecular calculations remains an open and challenging problem and is not the focus of this work.

ECPs thus form a middle ground enabling the study of relativistic effects of the core electrons on the valence ones by making relatively minor modifications of the existing scattering codes. At the same time, the use of ECPs leads to a valuable saving of computational time for the two-electron integral problem since the calculations of continuum spectra necessarily include large bases of continuum functions, which makes the integral calculations relatively more demanding compared to the bound-state problems of quantum chemistry.

In this work we describe the implementation of ECPs in the molecular R-matrix codes UKRmol+~\cite{jt772} and in the related interface QEC \cite{jt778} and the first applications to electron scattering from molecules containing heavy atoms. Section~\ref{sec:method} describes the implementation in detail. In Section~\ref{sec:results} we provide examples and discuss the performance of the calculations including ECPs. Finally, in Section~\ref{sec:conclusion} we discuss the limitations of the present approach and the future prospects for its extension.

\section{Methods}\label{sec:method}

In atomic units the molecular fixed-nuclei Hamiltonian for $n_{v}$ valence electrons with $n_{c}$ core electrons represented by ECPs has the following form
\begin{eqnarray}
    H &=& \sum_{i=1}^{n_{v}} h(i) + \sum_{j=1}^{n_{v}}\sum_{i>j} \frac{1}{\vert \mathbf{r}_{i} - \mathbf{r}_{j}\vert}+\sum_{j=1}^{nuc}\sum_{i>j} \frac{Z_{i}Z_{j}}{\vert \mathbf{R}_{i} - \mathbf{R}_{j}\vert},\\
    h(i) &=& -\frac{\Delta_{i}}{2}-\sum_{j=1}^{nuc}\frac{Z_{j}-n_{c}(j)}{\vert \mathbf{R}_{j}-\mathbf{r}_{i}\vert}+V_{ECP}^{j}(\mathbf{r}_{i}),
\end{eqnarray}
where $\mathbf{r}_{i}$ represent the spatial coordinates of the valence electrons and $\mathbf{R}_{i}$ are the positions of the atomic nuclei and $Z_{i}$ are their charges. The one-electron contribution $h(i)$ consists of the electron's kinetic energy and the nuclear attraction energy between the electron and each nucleus whose bare charge is effectively screened by the $n_{c}(j)$ core electrons represented by the ECP centred on that atom.

The ECP comprises a local term and a semi-local term which is further split between scalar and spin-orbit components
\begin{eqnarray}\label{eq:ecp}
    V_{ECP}(\mathbf{r}) = V_{lmax}(r) + \sum_{l=0}^{lmax}\sum_{m=-l}^{l}\vert X_{lm}\rangle (V_{l}(r)-V_{lmax}(r))\langle X_{lm}\vert + \sum_{l=0}^{lmax'}\sum_{m=-l}^{l}\Delta V_{l}(r)\vert X_{lm}\rangle \mathbf{s.l}\langle X_{lm}\vert,
\end{eqnarray}
where the semi-local terms contain the projectors on the real spherical harmonics $X_{lm}(\hat{\mathbf{r}})$. Here $\mathbf{r}$ and the angular-momentum projectors are taken with respect to the atom where the ECP is centred. The radial potentials are all parametrized by a linear combination of Gaussians multiplied by powers of the radial distance
\begin{eqnarray}
    V_{l}(r) = \sum_{j} c_{j}r^{m_{j}-2}\exp[-\gamma_{j}r^2].
\end{eqnarray}
For reasons of integrability the coefficients $m_{j}$ must not be smaller than zero.

\subsection{Implementing ECPs in UKRMol+}

ECPs can used within the standard workflow of the R-matrix method~\cite{jt772}. The implementation of ECPs in UKRmol+ is therefore reduced to evaluating the matrix elements of the ECP potential~\eqref{eq:ecp} for the basis of one-electron atom- and center-of-mass-centred functions representing the bound and the continuum electrons, respectively. The ECP integrals have been implemented in the GBTOlib library~\cite{gbtolib}, which calculates all molecular integrals required by UKRmol+. In GBTOlib the continuum can be represented using an arbitrary combination of GTOs and B-spline orbitals (BTOs) centred at the centre-of-mass~\cite{jt772}, while the bound electrons are represented using the standard atom-centred GTOs of quantum chemistry. Orthogonalization of the continuum orbitals against the set of orbitals representing the bound electrons then leads to mixing of all one-electron basis functions in the expansion of the molecular continuum orbitals. This in turn requires calculation of ECP matrix elements for all combinations of the one-electron basis functions.

Evaluation of ECP matrix elements between GTOs can be performed using known analytic expressions~\cite{shaw2017} or using highly optimized routines employed for the calculation of 2-electron hybrid integrals~\cite{jt772}. Therefore the pure GTO integrals will not be discussed here further. However, besides integrals over GTOs we require also integrals involving the numerical B-spline orbitals, if those are used to represent the continuum. These include 1-, 2- and 3-centre integrals between atom-centred GTOs and BTOs centred on the centre-of-mass. For that reason the formulae for the matrix elements will be presented first in the general form suitable for a numerical quadrature, regardless of the orbital type used. The general expressions are then specialized to the particular cases.

For completness we mention that both the semi-local and local terms involving BTOs and ECPs centred on the centre-of-mass are calculated directly in the real space trivially by means of radial quadratures.

\subsubsection{One electron basis functions}

The general form of the one-electron basis functions is
\begin{eqnarray}
    \phi_{lm}(\mathbf{r}) = R(r)X_{lm}(\hat{\mathbf{r}}),
\end{eqnarray}
where $R(r)$ is the radial part of the orbital.

For a contracted GTOs the expression specializes to~\cite{helgaker2013}
\begin{eqnarray}
    \phi_{lm}^{G}(\mathbf{r}) &=& N_{G} S_{lm}(\mathbf{r})\sum_{j=1}^{n_{c}}c_{j}\exp[-\alpha_{j}r^2] = N_{G} \sqrt{\frac{4\pi}{2l+1}}r^{l}\sum_{j=1}^{n_{c}}c_{j}\exp[-\alpha_{j}r^2] X_{lm}(\mathbf{r}),\\
    R_{G}(r) &=& N_{G} \sqrt{\frac{4\pi}{2l+1}}r^{l}\sum_{j=1}^{n_{c}}c_{j}\exp[-\alpha_{j}r^2],
\end{eqnarray}
where $N_{G}$ normalizes the orbital to unit charge density, $S_{lm}$ is the real solid harmonic, $c_{j}$ are the contraction coefficients and $\alpha_{j}$ are the contraction exponents.

In case of BTOs the radial part of the orbital is the numerical B-spline function. The resulting BTO as employed in GBTOlib is~\cite{jt772}
\begin{eqnarray}
    \phi_{lm}^{B_{i}}(\mathbf{r}) &=& N_{B_{i}} \frac{B_{i}(r)}{r}X_{lm}(\mathbf{r}),\\
    R_{B_i}(r)&=& N_{B_{i}} \frac{B_{i}(r)}{r},
\end{eqnarray}
where $N_{B_{i}}$ is again the normalization factor and $B_{i}(r)$ is the i-th B-spline function drawn from a basis of B-splines~\cite{bachau2001} covering a selected radial range. B-splines are piecewise polynomials with compact support which allow the construction of a highly accurate and linearly independent basis for the oscillating wavefunction of the unbound particle.

For clarity the normalization factors for the GTOs and BTOs will be omitted from the following expressions but are assumed to be included as overall multiplication factors.

\subsubsection{Translation of the BTO orbitals}

The atomic character of the ECPs and the angular-momentum projectors employed in the semi-local terms require that the integration over the semi-local component be performed most conveniently with respect to the atomic nucleus where the ECP is centred. This is also the approach used in the analytic evaluation of ECP integrals over GTOs and we use it for the evaluation of the mixed and pure BTO integrals too. When computing ECP integrals we encounter 1-, 2- and 3-centre integrals. 

Computing the GTO-only integrals makes use of the analytic form of the translation of the GTO with respect to an arbitrary centre. ECP integrals involving the numerical BTOs must be handled differently. One option is to compute the integrals in real space by rotating the coordinate system along the line connecting the centre-of-mass with the atomic nucleus allowing the analytic computation of the integral over the azimuthal angle  and the subsequent numerical  integration over the remaining two degrees of freedom. However, the finite support and the piecewise character of the radial B-splines complicates the angular integrals. While this is certainly a solvable problem we have implemented an alternative approach making use of the translation property of Fourier transforms which allows one to perform arbitrary translations easily in  momentum space.

Fourier transform of the BTO is simple
\begin{eqnarray}\label{eq:FT_full}
    \hat{\phi}^{B_{i}}_{lm}(\mathbf{k}) &=& \frac{1}{(2\pi)^{3/2}}\int d^{3}r \frac{B_{i}(r)}{r} X_{lm}(\hat{\mathbf{r}})\exp[\imath \mathbf{r}.\mathbf{k}]=\sqrt{\frac{2}{\pi}}\imath^{l}X_{lm}(\hat{\mathbf{k}})\frac{1}{k}\tau_{i,l}(k),\\
    \tau_{i,l}(k)&=&\int_{a_{i}}^{b_{i}} dr B_{i}(r)\hat{j}_{l}(kr),
\end{eqnarray}
and follows trivially from the partial wave expansion of the plane-wave in the basis of real spherical harmonics
\begin{eqnarray}
    \exp[\imath \mathbf{r}.\mathbf{k}] = 4\pi \sum_{l,m}\imath^{l}\frac{\hat{j}_{l}(kr)}{kr}X_{lm}(\hat{\mathbf{r}})X_{lm}(\hat{\mathbf{k}}).
\end{eqnarray}
We note that the Fourier transformation of the BTO preserves the angular momentum character of the orbital.

The factor $\tau_{i,l}(k)$ contains the Riccati-Bessel function $\hat{j}_{l}(kr)$ and is analogical to the spherical Bessel transform investigated earlier by Talman in the context of multi-centre integrals over numerical orbitals~\cite{talman2003,talman2009,koval2010}. An efficient computation of this factor is crucial, in particular for B-splines localized further away from the origin, where the integrand becomes highly oscillatory. The methods of Talman make use of logarithmic radial grids which are applicable to orbitals with an infinite range, such as Slater orbitals, but not to B-splines extending over the finite radial range $[a_{i}, b_{i}]$. Instead, to calculate $\tau_{i,l}(k)$ we implemented Levin quadrature~\cite{levin1996} which is tailored to integrands of this type for arbitrary values of $k$. 

The real-space representation of the BTO is obtained from the inverse Fourier transform
\begin{eqnarray}\label{eq:FT_BTO}
    \phi^{B_{i}}_{lm}(\mathbf{r}) &=& \frac{1}{(2\pi)^{3/2}}\int d^{3}k \hat{\phi}^{B_{i}}_{lm}(\mathbf{k}) \exp[-\imath \mathbf{k}.\mathbf{r}].
\end{eqnarray}
It follows that translation of the BTO to another centre $\mathbf{A}$ is performed by applying a phase shift to the momentum space coefficients
\begin{eqnarray}
    \phi^{B_{i}}_{lm}(\mathbf{r}-\mathbf{A}) &=& \frac{1}{(2\pi)^{3/2}}\int d^{3}k \hat{\phi}^{B_{i}}_{lm}(\mathbf{k}) \exp[-\imath \mathbf{k}.(\mathbf{r}-\mathbf{A})]\\
    &=&\frac{1}{(2\pi)^{3/2}}\int d^{3}k \left[\hat{\phi}^{B_{i}}_{lm}(\mathbf{k})\exp[\imath\mathbf{k}.\mathbf{A}]\right]\exp[-\imath \mathbf{k}.\mathbf{r}].
\end{eqnarray}
An equivalent relation can be obtained for any function equipped with a Fourier transform. In particular, we will use it for the BTO-only integrals to translate the local component of the ECP to the centre-of-mass.

To perform the ECP integrals over BTOs we  need to evaluate, at a given distance $r_{A}$ from the atom, a projection of the BTO onto spherical harmonics centred on the atom, $\mathbf{A}$. This projection is straightforwardly derived using the Fourier representation~\eqref{eq:FT_BTO} and the formulae listed above:
\begin{eqnarray}
    \langle X_{l'm'}(\hat{\mathbf{r}}_{A})\vert\phi_{lm}^{B_{i}}\rangle\big\vert_{r_{A}} &=& \frac{1}{(2\pi)^{3/2}}\int d^{3}k \hat{\phi}^{B_{i}}_{lm}(\mathbf{k}) \int d\Omega_{\mathbf{r}_{A}} X_{l'm'}(\hat{\mathbf{r}}_{A})\exp[-\imath \mathbf{k}.\mathbf{r}]\nonumber\\
    &=& \frac{1}{(2\pi)^{3/2}}\int d^{3}k \hat{\phi}^{B_{i}}_{lm}(\mathbf{k}) \int d\Omega_{\mathbf{r}_{A}} X_{l'm'}(\hat{\mathbf{r}}_{A})\exp[-\imath \mathbf{k}.\hat{\mathbf{r}}_{A}]\exp[-\imath\mathbf{k}.\mathbf{A}]\nonumber\\
    &=&\frac{4\pi}{(2\pi)^{3/2}}\imath^{l'}\int d^{3}k \hat{\phi}^{B_{i}}_{lm}(\mathbf{k})\frac{\hat{j}_{l'}(k r_{A})}{kr_{A}}X_{l'm'}(\hat{\mathbf{k}})\exp[-\imath\mathbf{k}.\mathbf{A}]\nonumber\\
    &=&8\sum_{\lambda,\mu}\langle lm\vert l'm'\vert\lambda\mu\rangle_{R}X_{\lambda\mu}(\hat{\mathbf{A}})\imath^{\lambda+l'+l}\int_{0}^{\infty} dk k \tau_{i,l}(k)\frac{\hat{j}_{l'}(k r_{A})}{kr_{A}}\frac{\hat{j}_{\lambda}(k A)}{kA}.\label{eq:BTO_proj}
\end{eqnarray}
The partial wave expansion and the expression~\eqref{eq:FT_full} were used to simplify the derivation. Here the symbol $\langle lm\vert l'm'\vert\lambda\mu\rangle_{R}$ stands for the Gaunt coefficient for real spherical harmonics~\cite{homeier1996}.

\subsection{Semi-local term}

The general expression for the matrix element of the semi-local term can be written as a radial integral with respect to the atomic centre $\mathbf{A}$
\begin{eqnarray}
    \langle \phi_{i}\vert \hat{V}^{s-l}\vert\phi_{j}\rangle &=& \sum_{l=0}^{lmax}\sum_{m=-l}^{l}\int_{0}^{\infty} dr_{A} r_{A}^{2} \langle\phi_{i}\vert X_{lm}(\hat{\mathbf{r}}_{A})\rangle V_{l}^{s-l}(r_{A})\langle X_{lm}(\hat{\mathbf{r}}_{A})\vert\phi_{j}\rangle,\\
    V_{l}^{s-l}(r_{A}) &=& V_{l}(r_{A})-V_{lmax}(r_{A}),
\end{eqnarray}
where the two functions $\phi_i$ and $\phi_j$ stand for any type of basis function. Computing the integral involving the semi-local potential requires calculation of the projections of the basis functions on the atom-centred spherical harmonics. When one of the functions is a BTO we employ the formula~\eqref{eq:BTO_proj}. For hybrid integrals the GTO partial waves are obtained from a plane-wave expansion. Here, we provide the working formulae for the particular classes of integrals involving BTOs.

\subsubsection{BTO - BTO class} 

The expression for this integral is arrived at by employing the angular projections for the BTO orbitals in the momentum-space representation~\eqref{eq:BTO_proj} and integrating the result over the radial coordinate $r_{A}$ centred on the ECP
\begin{eqnarray}
    \langle\phi_{l_{i}m_{i}}^{B_{i}} \vert \hat{V}_{s-l}\vert \phi_{l_{j}m_{j}}^{B_{j}}\rangle &=& 64\sum_{l,m}\sum_{\lambda_{i},\mu_{i}}\langle l_{i}m_{i}\vert lm\vert\lambda_{i}\mu_{i}\rangle_{R} X_{\lambda_{i}\mu_{i}}(\hat{\mathbf{A}})\nonumber\\
    &\times&\sum_{\lambda_{j},\mu_{j}}\langle l_{j}m_{j}\vert lm\vert\lambda_{j}\mu_{j}\rangle_{R} X_{\lambda_{j}\mu_{j}}(\hat{\mathbf{A}})(-1)^{\frac{l_{j}-l_{i}+\lambda_{2}+\lambda_{1}}{2}}\nonumber\\
    &\times&\int_{0}^{\infty} dk_{1}\int_{0}^{\infty} dk_{2} k_{1}k_{2} \frac{\hat{j}_{\lambda_{i}}(k_{1} A)}{k_{1}A}\frac{\hat{j}_{\lambda_{j}}(k_{2} A)}{k_{2}A}\tau_{i,l_{i}}(k_{1})\tau_{j,l_{j}}(k_{2})\alpha_{l}(k_{1},k_{2}),\\
    \alpha_{l}(k_{1},k_{2}) &=& \int dr_{A} r_{A}^{2}\frac{\hat{j}_{l}(k_{1} r_{A})}{k_{1}r_{A}}\frac{\hat{j}_{l}(k_{2} r_{A})}{k_{2}r_{A}}V_{l}^{s-l}(r_{A}) .
\end{eqnarray}

The momentum-space integrals are currently computed using sufficiently large momentum-space quadratures. The BTOs located farther away from the centre-of-mass lead to a highly oscillatory momentum-space representation $\tau_{i,l}(k)$, but in many cases the heavy atoms containing ECPs are localized close to the centre so that they do not overlap with those BTOs in the real space and the corresponding ECP integral is zero. Although dense and wide momentum-space grids are required, the computation is still manageable and provides orders of magnitude computational savings compared to the all-electron calculation. Nevertheless, the development of more optimal quadratures suitable for these oscillatory integrals is highly desirable and will be the subject of follow-up work.

\subsubsection{GTO - BTO class}

To evaluate this class we need to combine the numerically evaluated angular projection of the GTO in real space with the angular projection on the BTO evaluated using the Fourier representation. The result is
\begin{eqnarray}
    \langle\phi_{l_{i}m_{i}}^{G_{i}} \vert \hat{V}_{s-l}\vert \phi_{l_{j}m_{j}}^{B_{j}}\rangle &=& 8 \sum_{l,m}\sum_{\lambda\mu}\langle lm\vert l_{j}m_{j}\vert\lambda\mu\rangle_{R}X_{\lambda\mu}(\mathbf{A})(-1)^{\frac{l-l_{j}+\lambda}{2}}\nonumber\\
    &\times&\int dk \frac{\hat{j}_{\lambda}(kA)}{kA}\tau_{j,l_{j}}(k)\rho_{lm}(k),\\
    \rho_{lm}(k)&=&\int_{0}^{\infty} dr_{A} r_{A}^{2} \frac{\hat{j}_{\lambda}(kr_{A})}{kr_{A}} \langle\phi_{l_{i}m_{i}}^{G_{i}} \vert X(\hat{\mathbf{r}}_{A})\rangle\big\vert_{r_{A}}V_{l}^{s-l}(r_{A}).
\end{eqnarray}

\subsection{Local term}

\begin{eqnarray}
    \langle \phi_{i}\vert \hat{V}^{loc}\vert\phi_{j}\rangle &=& \int d^{3}r_{A} \phi_{i}(\mathbf{r}_{A}) V_{l}^{loc}(r_{A})\phi_{j}(\mathbf{r}_{A}),\\
    V^{loc}(r_{A}) &=& V_{lmax}(r_{A}).
\end{eqnarray}

This type of integral can be formally recast in the form of a projection of the product of the two basis functions on the spherical harmonic $X_{00}$ followed by integration with respect to the ECP centre
\begin{eqnarray}
    \int d^{3}r_{A} \phi_{i}(\mathbf{r}_{A}) V_{l}^{loc}(r_{A})\phi_{j}(\mathbf{r}_{A}) = \sqrt{4\pi}\int dr_{A} r_{A}^{2} V_{l}^{loc}(r_{A}) \langle X_{00}\vert\phi_{i}(\mathbf{r}_{A})\phi_{j}(\mathbf{r}_{A})\rangle_{r_{A}}.
\end{eqnarray}

\subsubsection{BTO - BTO class}

The pure BTO case can be computed fairly straightforwardly by representing the local potential in the momentum space followed by its translation to the centre-of-mass while keeping the BTOs in their real-space representation
\begin{eqnarray}
    \langle \phi_{l_{i}m_{i}}^{B_{i}}\vert \hat{V}^{loc}\vert\phi_{l_{j}m_{j}}^{B_{j}}\rangle &=& \frac{2}{\pi}\sum_{\lambda\mu}\langle l_{1}m_{1}\vert l_{2}m_{2}\vert \lambda\mu\rangle_{R}X_{\lambda\mu}(\mathbf{A})\int_{0}^{\infty}dk k^2 \frac{\hat{j}_{\lambda}(kA)}{kA} \tau_{i,j,\lambda}(k)\epsilon_{0}(k),\\
    \epsilon_{0}(k) &=& \int_{0}^{\infty} r^2 \frac{\hat{j}_{0}(kr)}{kr} V_{lmax}(r),\\
    \tau_{i,j,\lambda} &=&\int_{0}^{\infty} \frac{\hat{j}_{\lambda}(kr)}{kr}B_{i}(r)B_{j}(r).
\end{eqnarray}


\subsubsection{GTO - BTO class}

In this case one is tempted to combine the terms from the local potential ($r^{n}\exp[-\gamma r^2]$) with the atom-centred Gaussian to obtain overlap-type integral between BTOs and GTOs. Those integrals can be handled very efficiently using GBTOlib. However, the radial parts of the local potential may contain odd powers of the radial coordinate preventing the polynomial-type translation of the prefactor to the centre-of-mass. This makes the BTO-GTO class for general local potentials less amenable to analytic techniques. Instead, this type of integral can be handled by a direct 3D quadrature with respect to the centre of the BTO and will be implemented in future releases of GBTOlib.

\subsection{QEC implementation}

Quantemol electron collisions (QEC) \cite{jt778} is an expert system that performs electron-molecule scattering calculations, using both the UKRMol+ codes  \cite{jt772} and the MOLPRO quantum chemistry package \cite{werner2012,werner2020}. Although QEC has mainly been applied to systems containing fairly light elements \cite{snoeckx2023, zhang2023, sapunar2024}, it has made some use of ECPs in the past. Originally, ECPs were implemented in QEC to calculate electron-impact ionization cross sections\cite{jt823} using the binary encounter Bethe (BEB) method of Kim and Rudd \cite{kim1994}. This was straightforward since BEB calculations require input from MOLPRO only, and are independent of the R-matrix calculation. The use of ECPs within the R-matrix method now enables QEC to calculate the full set of cross sections for molecules containing heavy elements.

QEC uses the ECPs provided by the Stuttgart/K\"{o}ln group \cite{nicklass1995}. These potentials are named ECP$n$XYZ, where $n$ is the number of core electrons represented by the ECP, and the XYZ label indicates the level of theory used when constructing the pseudovalence orbitals. The X label indicates the reference system used to generate the ECP, X=S if a single-valence-electron ion was used, and X=M if a neutral atom was used. The YZ label denotes the method used, YZ=HF for the non-relativistic Hartree-Fock method, YZ=WB for quasi-relativistic Wood-Boring method, and YZ=DF for the relativistic Dirac-Fock method. The calculations reported in this work use semi-relativistic ECPs labelled ECP$n$MWB.

A range of valence-electron basis sets can be used with these potentials. Associated basis sets named ECP$n$MWB \cite{ECPn} are provided in MOLPRO. In some cases these are double zeta bases, though for some elements a range of triple and quadruple zeta basis sets are available, as well a basis sets including diffuse functions. The Karlsruhe (named def2-XYZ) basis sets \cite{weigend2005} are also compatible with the Stuttgart/K\"{o}ln ECPs for some heavy elements.

\section{Results and discussion}\label{sec:results}
In this section we provide a set of electron scattering cross sections for a number of molecular targets. In Section \ref{sec:br2} we provide a proof-of-principle test case using the bromine molecule, for which calculations using ECPs may be benchmarked against an all-electron calculation. Electron scattering cross sections are then presented in Sections \ref{sec:sibr4} and \ref{sec:wh} for the heavier \sibr\ and WH systems, that are beyond the reach of all-electron calculations. 

The ECP capability has been implemented for photoionization calculations too. In Section~\ref{sec:ch3i} we present one example for valence ionization of the CH$_3$I molecule.

\subsection{Br$_2$}\label{sec:br2}

The Br$_2$ molecule serves as a proof-of-principle test case, in which cross sections can be calculated using ECPs and valence basis sets, as well as all-electron basis sets. We compare cross sections for Br$_2$ calculated using the all-electron cc-pVTZ basis set with those obtained using a 28-electron scalar-relativistic core potential (ECP28MWB) for each Br atom, with the remaining 7 valence electrons of each Br treated using the ECP28MWB\_VTZ basis set. 
In both cases the unbound electron was treated using a basis of Gaussian-type orbitals (GTOs) with angular momentum $l \leq 4$. The $R$-matrix radius was set to 14 Bohr.
All cross sections were calculated using the close-coupling method. The calculations used the irreducible representations of the $D_{2h}$ point group. Using these irreducible representations, the ground state configuration of Br$_2$ may be expressed as $(1-9a_g)^2(1-4b_{3u})^2(1-4b_{2u})^2 1b_{1g}^2,(1-8b_{1u})^2 (1-4b_{2g})^2 (1-4b_{3g})^2 1a_u^{2}$. The calculation using the cc-pVTZ basis set used a target active space in which 62 electrons were frozen, $(1-9a_g,1-3b_{3u},1-3b_{2u},1b_{1g},1-8b_{1u},1-3b_{2g},1-3b_{3g},1a_u)^{62}$, leaving 8 valence active electrons distributed among 8 active orbitals, $(10-11a_g,4b_{3u},4b_{2u},9-10b_{1u},4b_{2g},4b_{3g})^{8}$. The active space settings for the calculations using ECPs mirror this calculation: 56 electrons are handled using the ECPs, leaving a residual target with 14-electron, 6 of which are kept frozen, occupying the lowest $a_g$ and $b_{1u}$ orbitals of the valence-electron basis set, $(1-2a_g, 1b_{1u})^6$, and the remaining 8 electrons are allowed to occupy the active orbitals $(3-4a_g, 1b_{3u}, 1b_{2u}, 2-3b_{1u}, 1b_{2g}, 1b_{3g})^{8}$. Six additional virtual orbitals were also included in both calculations.

The equilibrium geometry used for Br$_2$ assumed an internuclear separation of 2.2756 \AA. 

\begin{figure}[t]
    \centering
    \includegraphics[width=0.5\linewidth]{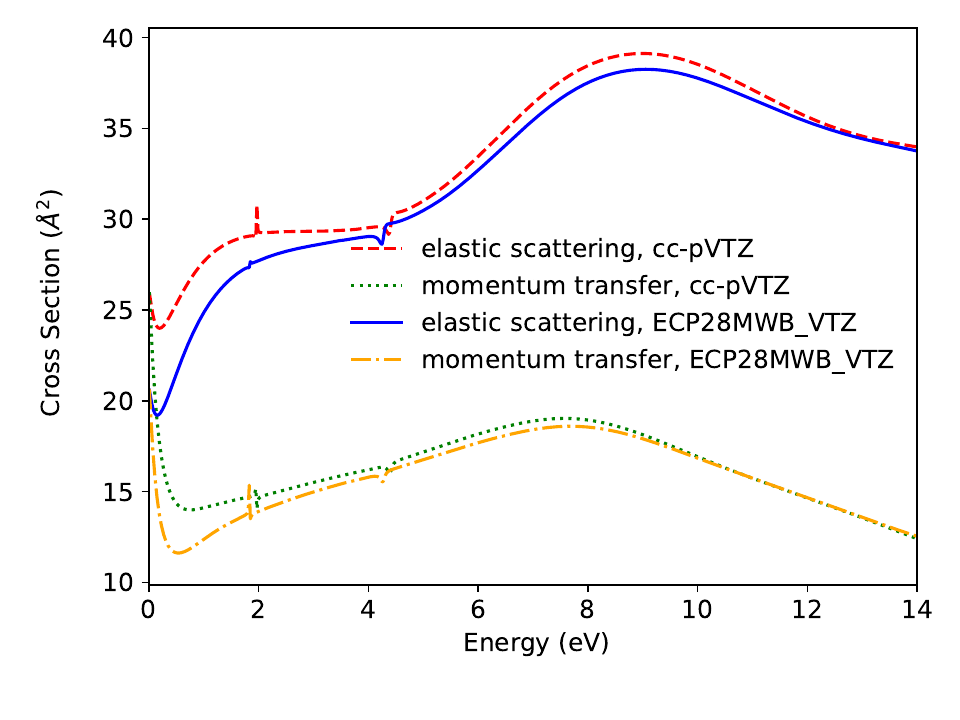}
    \caption{Elastic scattering and momentum transfer cross sections for Br$_2$.}
    \label{br2ecs}
\end{figure}

Figure \ref{br2ecs} shows the elastic scattering and momentum transfer cross sections for Br$_2$, calculated with and without the use of ECPs. A strong level of agreement is observed in both cross sections, including the positions of some small resonance features. Small differences in the cross section values are apparent at low energies. It is tempting to ascribe these differences to the semi-relativistic nature of the ECP and valence basis set. However, it is not clear if this is the case, since a non-relativistic ECP for Br, that would provide definitive proof of such an effect, is not available.  

\begin{figure}[t]
    \centering
    \includegraphics[width=0.45\linewidth]{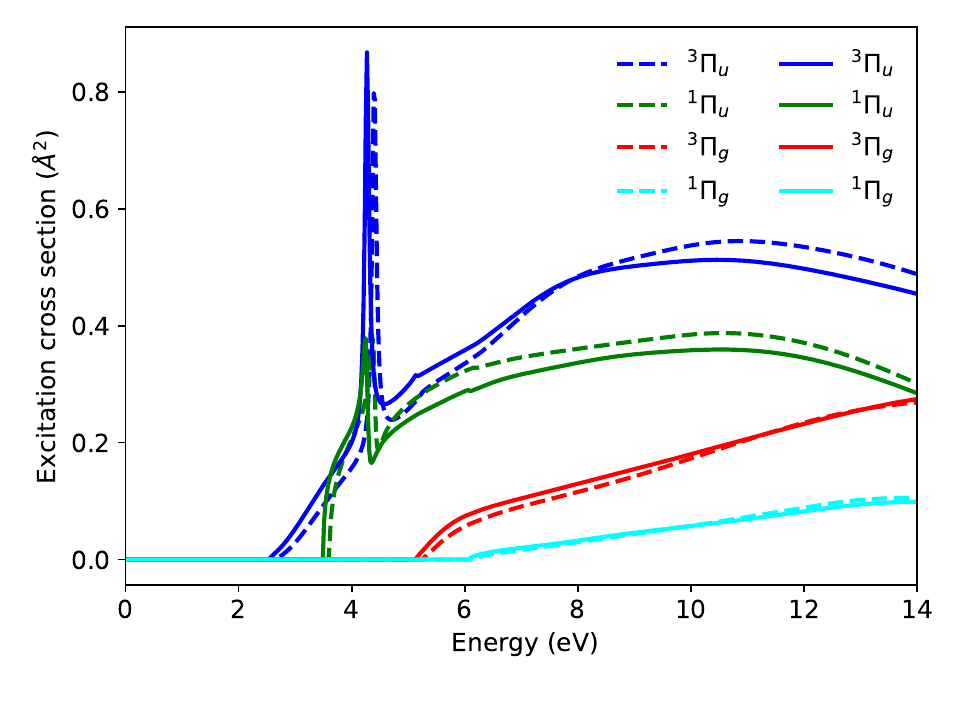}
    \includegraphics[width=0.45\linewidth]{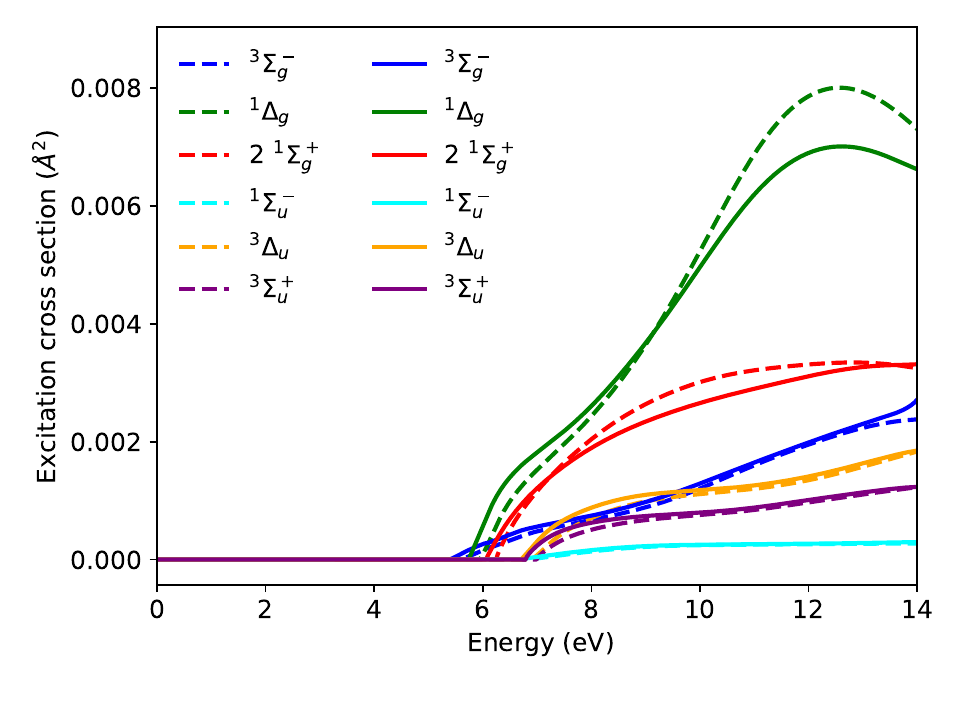}
    \caption{Electronic excitation cross sections for Br$_2$ $\Pi$ states (left) and $\Sigma$ and $\Delta$ states (right). Dashed lines are the cross sections obtained using the all-electron cc-pVTZ basis set, solid lines are the cross sections obtained using an ECP.}
    \label{br2ex}
\end{figure}

Figure \ref{br2ex} shows the calculated electronic excitation cross sections for Br$_2$. Again, very good agreement is observed between the two results, including the resonant features of the $1\ ^3\Pi_u$ state above 4 eV, as well as a similar feature seen for excitation to the $1\ ^1\Pi_u$ state. This also extends to the vertical excitation energies, given in Table \ref{brvee}. The largest relative error in the excitation energies for the first few excited states was found to be 4\%. The excitation energies are also in good agreement with recent non-relativistic calculations \cite{vadhel2024}. The main difference found is that the $^1\Pi_g$ and $2\ ^1\Sigma_g^+$ states are almost degenerate in the calculation using ECPs, while these states are distinctly separated in the non-relativistic all-electron calculations given here  and in Ref.~\cite{vadhel2024}.  

\begin{table}[h]
 \caption{Comparison of vertical excitation energies for Br$_2$ calculated using all-electron basis sets with those obtained using a 28-electron ECP for each Br.}
\begin{tabular}{cccc}
State & \multicolumn{3}{c}{Excitation energy (eV)}  \\
\hline 
  &  cc-pVTZ  & ECP &   Ref.~\cite{vadhel2024}\\
 \hline
    1~$^3\Pi_u$ & 2.67 & 2.54 & 2.64 \\
    1~$^1\Pi_u$ & 3.61 & 3.50 & 3.56\\
    1~$^3\Pi_g$ & 5.26 & 5.14 & 5.22\\
    1~$^3\Sigma_g^-$ & 5.59 & 5.38 & 5.71\\
    1~$^1\Delta_g$ &  5.95  & 5.75 & 6.07 \\
    1~$^1\Pi_g$ & 6.17 & 6.07 & 6.10\\
    2~$^1\Sigma_g^+$ & 6.26 & 6.07 & 6.39 \\
    1~$^1\Sigma_u^-$ & 6.82 & 6.62 & 6.91 \\
    1~$^3\Delta_u$ &   6.92 & 6.72 & 6.99 \\
    1~$^3\Sigma_u^+$ & 7.00 & 6.80 & 7.10 \\
    \hline
\end{tabular}
\label{brvee}
\end{table}

\begin{figure}[h]
    \centering
    \includegraphics[width=0.5\linewidth]{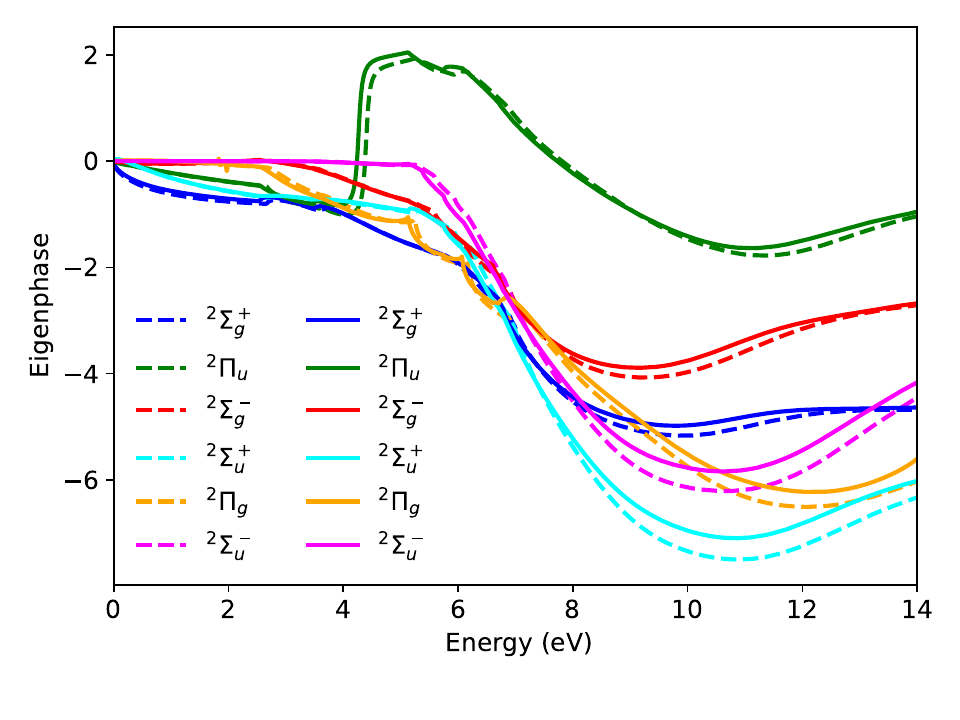}
    \caption{Eigenphases for Br$_2$. Solid lines are the results using the 28-electron ECP for each Br atom, dashed lines are the results from the all-electron calculation.}
    \label{br2eigs}
\end{figure}

In addition to the cross section data, perhaps a more detailed and fundamental test of the similarity in the two approaches may be made by comparing the calculated eigenphases in each case. Figure \ref{br2eigs} shows the calculated eigenphases for Br$_2$. Excellent agreement is observed between the eigenphases obtained using the all-electron cc-pVTZ basis set (dashed lines) and those obtained using ECPs (solid lines). The good agreement seen here confirms that both calculations capture resonant features and electronic excitation processes at almost identical energies.

\subsection{SiBr$_4$}\label{sec:sibr4}


Following our proof-of-principle calculations of Br$_2$ cross section data, we calculate cross sections for silicon tetrabromide, a compound used as a precursor in deposition of silicon nitride. Experimental data on \sibr\ is relatively scarce, though measurements of resonance energies and dissociative electron attachment ion yields are available \cite{wan1989, omarsson2014}. Previous calculations of \sibr\ cross sections may be found in Refs.\cite{varella1999} and \cite{bettega2011}, where cross sections were calculated using the Schwinger multichannel method with pseudopotentials at both the static exchange (SE) and static exchange plus polarization (SEP) levels.

\sibr\ is a 154-electron system, and although all-electron basis sets could be used for both Si and Br, an all-electron calculation would be impractical on computational grounds. Both the calculation runtime and memory requirements of the R-matrix method scale strongly with the number of electrons in the target molecule. Our calculations used a 28-electron scalar-relativistic core potential (ECP28MWB) for each Br atom, and the remaining 7 valence electrons of each Br were treated using the ECP28MWB\_VTZ basis set. A cc-pVTZ basis set was used for the Si atom. These settings reduce the number of electrons to be treated using basis sets from 154 to 42. A pure GTO continuum basis with maximum angular momentum $l_{\rm max}=4$ was used to represent the unbound electron, and the $R$-matrix radius was set to 14 Bohr.
The equilibrium geometry used for \sibr\ is given in Table \ref{sibrgeom}. Each of the Si-Br bond lengths is 2.2152 \AA. The tetrahedral \sibr\ molecule belongs to the $T_d$ point group in its equilibrium geometry, and our calculations are carried out using the irreducible representations of the $C_{2v}$ point group, the highest Abelian subgroup of $T_d$.
\begin{table}[h]
    \centering
    \begin{tabular}{c|c|c|c|}
      Atom & x(\AA) & y(\AA)  & z (\AA) \\
      \hline 
       Si  &  0     &   0     &   0 \\
       Br  &  0     & -1.8087 &  1.2790 \\
       Br  &  0     &  1.8087 &  1.2790 \\
       Br  & 1.8087 &   0     & -1.2790 \\
       Br  &-1.8087 &   0     & -1.2790 \\
    \end{tabular}
    \caption{Equilibrium geometry of \sibr\ in the centre-of-mass frame.}
    \label{sibrgeom}
\end{table}

\begin{figure}
    \centering
    \includegraphics[width=0.5\linewidth]{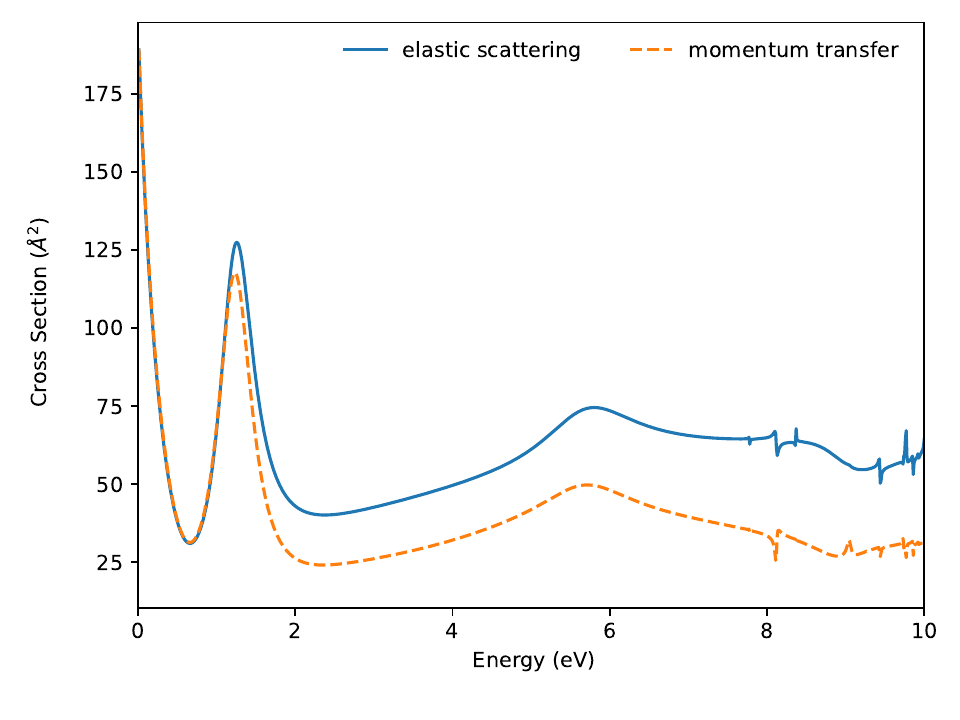}
    \caption{Elastic scattering and momentum transfer cross sections for \sibr.}
    \label{sibr4ecs}
\end{figure}

Figure \ref{sibr4ecs} shows the calculated elastic scattering and momentum transfer cross sections for \sibr, calculated using the SEP method. Prominent resonances appear at around 1.2 eV and 6 eV, in good agreement with electron transmission measurements \cite{wan1989, omarsson2014}, as well as calculations using the Schwinger multichannel method \cite{bettega2011}. A symmetry analysis of these resonances shows that they are of $T_2$ and $E$ symmetry, as also shown in Ref.\cite{bettega2011}. In the SEP calculations, pseudoresonances begin to appear just below 8 eV due to the neglect of electronic excitation processes in the SEP method.

\begin{figure}
    \includegraphics[width=0.5\linewidth]{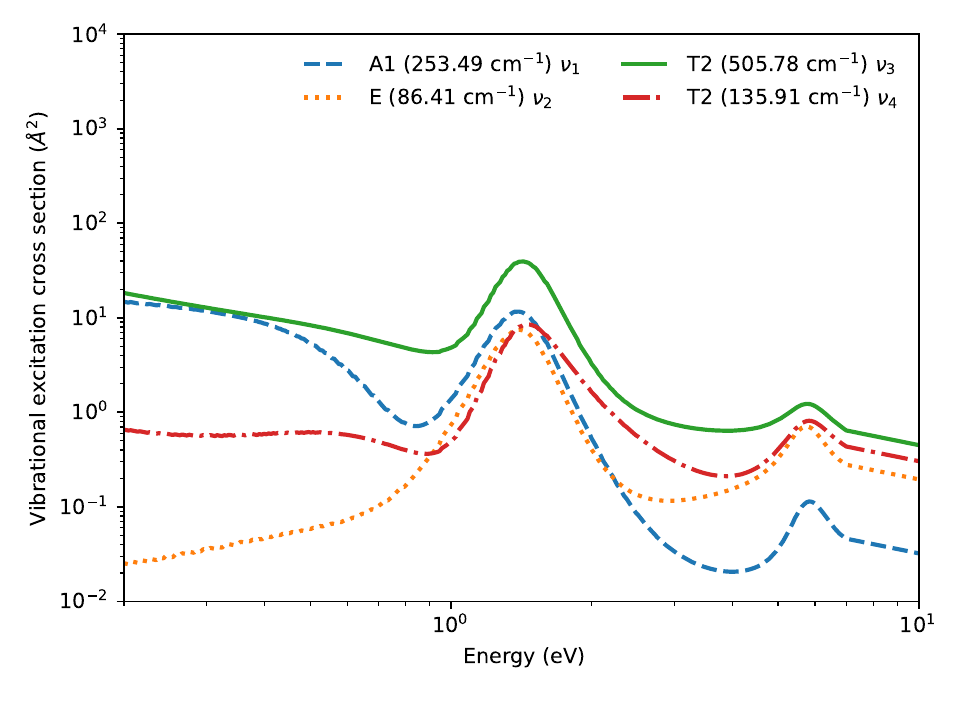}
    \caption{Vibrational excitation cross sections for \sibr.}
    \label{sibr4ve}
\end{figure}

\begin{table}[ht]
 \caption{Comparison of vibrational frequencies for \sibr\ calculated using QEC with data from NIST \cite{nist};
 $\Gamma$ is the symmetry group of mode in the T$_d$ point group.
} 
    \centering
    \begin{tabular}{lllrr}
    \hline
      label& mode  &$\Gamma$& \multicolumn{2}{c}{frequency (cm$^{-1}$)} \\
       \hline
            & & & NIST & QEC \\
       \hline
        $\nu_1$ & symmetric stretch& A$_1$   & 249  & 253.49 \\
        $\nu_2$ &  bending deformation& E    &  90  &  86.41 \\
        $\nu_3$ & asymmetric stretching& T$_2$    & 487  & 505.78 \\
        $\nu_4$ & bend& T$_2$   & 137  & 135.91 \\
        \hline
    \end{tabular}
    \label{tab:sibr4vf}
\end{table}
Table \ref{tab:sibr4vf} gives vibrational frequencies calculated by MOLPRO; these agree with values from NIST \cite{nist} to within 5\% for all modes.
Figure \ref{sibr4ve} shows the vibrational excitation cross sections for $v=0\rightarrow 1$, calculated using the method of Kokoouline {\it et al.} \cite{kokoouline2018}. The dipole-allowed excitation of the T$_2$ asymmetric stretching mode is strongest at all energies, though significant contributions appear from all modes, including the symmetric stretch and bending deformation modes that are not dipole-allowed.

\begin{figure}
    \centering
    \includegraphics[width=0.5\linewidth]{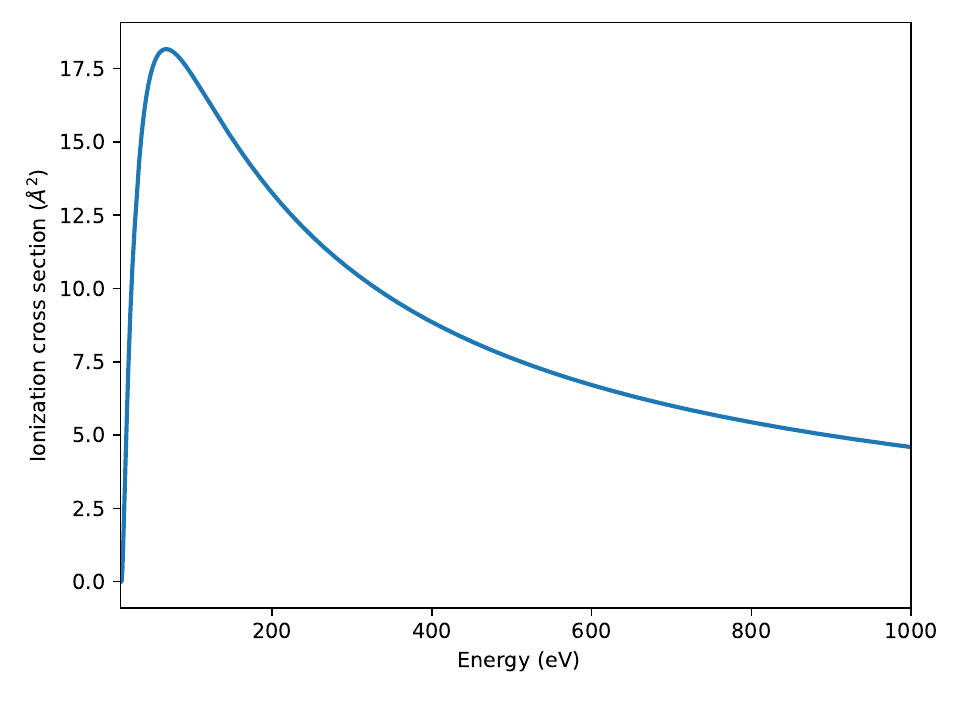}
    \caption{Total ionization cross section for \sibr.}
    \label{sibr4i}
\end{figure}

Figure \ref{sibr4i} shows the total ionization cross section for \sibr, calculated using the Binary-Encounter Bethe method \cite{kim1994}. The ionization potential obtained using Koopman's theorem was found to be 11.59 eV, in reasonable agreement with the measured value of 10.62 eV \cite{nist}.

\subsection{WH}\label{sec:wh}


Progressing to even heavier systems, a set of scattering cross sections for WH were calculated. 
Tungsten (W) is a preferred plasma-facing material in experimental fusion plasma including the ITER. Emission from the WH (in practice WD)
electronic band $^6\Pi \rightarrow ^6\Sigma^+$ have
have been observed in such plasma environments \cite{19BrPoSe.WD}. These emissions provide the easiest way to monitor W errosion and
molecule formation with electron collisions providing the main mechanism of populating the $^6\Pi$ state. Electronic excitation
cross for this species are therefore important.

For WH (WD), the cc-pVTZ basis set was used for the H atom, and a 60-electron scalar-relativistic core potential (ECP60MWB) was used for the W atom, leaving the remaining 14 valence electrons of W to be treated with a basis set. 
In this case, the choice of basis set required careful consideration of the angular momentum components retained in the basis. The basis set that accompanies the ECP60MWB core potential for W contains $g$ functions, and could only be adequately contained with an inner region sphere of around 20 Bohr. This in turn affected the choice of continuum basis set. Gaussian-type orbitals centred on the centre-of-mass are suitably accurate for inner-region radii of up to 14 Bohr. For larger inner regions, B-spline-type orbitals (BTOs) are available in the $R$-matrix codes \cite{jt772}, although this feature is not available in QEC.

In order to perform a calculation using the standard parameters of the available QEC package (a purely GTO continuum basis and inner region radius of up to 14 Bohr), alternative target basis sets were considered. In particular, the Karlsruhe family of basis sets was considered, as these can be used in conjunction with the Stuttgart ECPs for some elements \cite{weigend2005}. The def2-TZVP basis set was chosen, as it contains up to and including $f$ basis functions. Using this basis, calculations could be completed using an inner region radius of 14 Bohr. Given the numerical issues encountered when using the Stuttgart basis set, it was thought prudent to test the sensitivity of cross sections obtained using the def2-TZVP basis set to the inner region radius. 

An additional calculation was carried out using an inner region sphere of 24 Bohr. These calculations used BTOs of order 4 starting at a radius of 11 Bohr. One of the main computational demands of such calculations is the evaluation of mixed GTO/BTO integrals. These integrals were calculated using Gauss-Legendre quadrature with stepsize $\Delta r$ = 0.25, in which the Coulomb potential was represented by Legendre expansion with a maximum angular momentum 8. 

The equilibrium bond length calculated by MOLPRO was found to be 1.8377~\AA. All cross sections were calculated using the close-coupling method, since the first excitation threshold was found to be close to 1 eV. SEP calculations are likely to display pseudoresonances starting at energies close to this threshold, leaving a physically meaningful cross section only at very low energies.  The active space for the close-coupling calculations allowed 7 active electrons to occupy 10 orbitals. The active space consisted of 8 frozen electrons in the orbitals $(1-2a_1, 1b_1, 1b_2)^8$, and the 7 active electrons occupying the orbitals $(3-5a_1, 2-3b_1, 2-3b_2, 1a_2)$. Two additional virtual orbitals were also included.

\begin{figure}[t]
    \centering
    \includegraphics[width=0.5\linewidth]
    {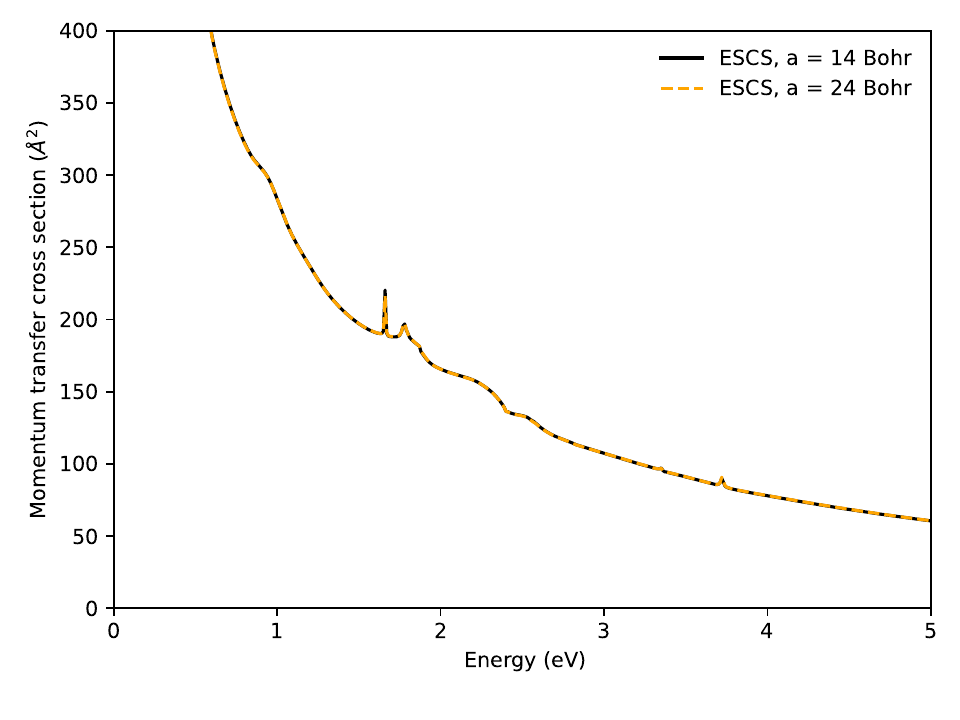}
    \caption{Elastic scattering cross section for WH using different inner-region sphere radii. The calculation using a sphere radius of 24 Bohr used B-spline continuum functions (see text for details).}
    \label{whecs}
\end{figure}

Figure \ref{whecs} shows the elastic scattering cross section for WH, calculated using inner-region sphere sizes of 14 Bohr and a GTO continuum basis, as well as a calculation using an inner region radius of 24 Bohr, and a mixed GTO/BTO continuum basis. In both cases, the continuum basis retained a maximum angular momentum of $l=4$. Since WH is a highly polar molecule, the dipole Born correction was added to the cross section to account for higher angular momenta. This correction increases the cross section significantly at low energies. A series of prominent resonances are visible between energies of 2 eV and 4 eV, indicating the likely presence of excited target states in this energy range. As can be seen in Fig. \ref{whecs}, very little sensitivity to sphere size was observed in the cross sections.

The vertical excitation energies calculated using QEC may be compared with those given in Ref. \cite{ma1991}. As shown in Table \ref{wh_vee}, we find a significant number of additional excited states that were not listed in Ref.~\cite{ma1991}, particularly doublet and quartet states. The level of agreement between our calculations and those given in Ref.~\cite{ma1991} is generally very good, with perhaps the exception of the $1\ ^6\Pi$ state, which appears in our calculation at a significantly higher energy than was found in Ref.~\cite{ma1991}. Additionally, the energetic order of the $2\ ^4\Pi$, $2\ ^4\Delta$, and $1\ ^6\Delta$ states differs in the two calculations, although these states lie with a narrow energy range of less than 0.5 eV in both cases.

\begin{table}
\caption{Excitation energies for electronic states of WH calculated using QEC compared to those of Ref. \cite{ma1991}.}
    \centering
    \begin{tabular}{lcc}
    State & \multicolumn{2}{c}{Excitation energy (eV)} \\
    \hline
     & this work & Ref. \cite{ma1991} \\
    \hline
    1 $^4\Delta$    & 0.855     & 0.749    \\
    1 $^4\Pi$       & 1.046     & 0.954    \\
    1 $^4\Sigma^+$  & 1.875     & 1.829\\
    1 $^2\Sigma^+$  & 2.173     &  -   \\
    1 $^2\Delta$    & 2.207     & 1.947\\
    2 $^2\Delta$    & 2.221     & 2.070\\
    1 $^6\Pi$       & 2.233     & 1.515\\
    1 $^2\Pi$       & 2.243     &  -   \\
    1 $^2\Sigma^-$  & 2.315     &  -   \\
    2 $^4\Pi$       & 2.321     & 2.555    \\
    2 $^4\Delta$    & 2.402     & 2.244    \\
    2 $^2\Pi$       & 2.409     &  -   \\
    3 $^2\Pi$       & 2.452     &  -   \\
    3 $^4\Delta$    & 2.457     & -    \\
    4 $^2\Pi$       & 2.464     &  -   \\
    3 $^2\Delta$    & 2.464     & 2.072\\
    1 $^6\Delta$    & 2.495     & 2.099 \\
    1 $^4\Sigma^-$  & 2.514     &  - \\
    3 $^4\Pi$       & 2.557     &  -    \\
    2 $^4\Sigma^+$  & 2.836     &  - \\
    4 $^4\Pi$       & 3.110     &  -    \\
    2 $^6\Sigma^+$  & 3.354     & 2.772 \\
    5 $^4\Pi$       & 3.402     &  -    \\
    2 $^4\Sigma^-$  & 3.424     &  - \\
    6 $^4\Pi$       & 3.500     &  -    \\
    7 $^4\Pi$       & 3.599     &  -    \\
    2 $^6\Pi$       & 3.759     & 4.223     \\
      \hline
    \end{tabular}
    \label{wh_vee}
\end{table}

\begin{figure}
    \centering
    \includegraphics[width=0.5\linewidth]{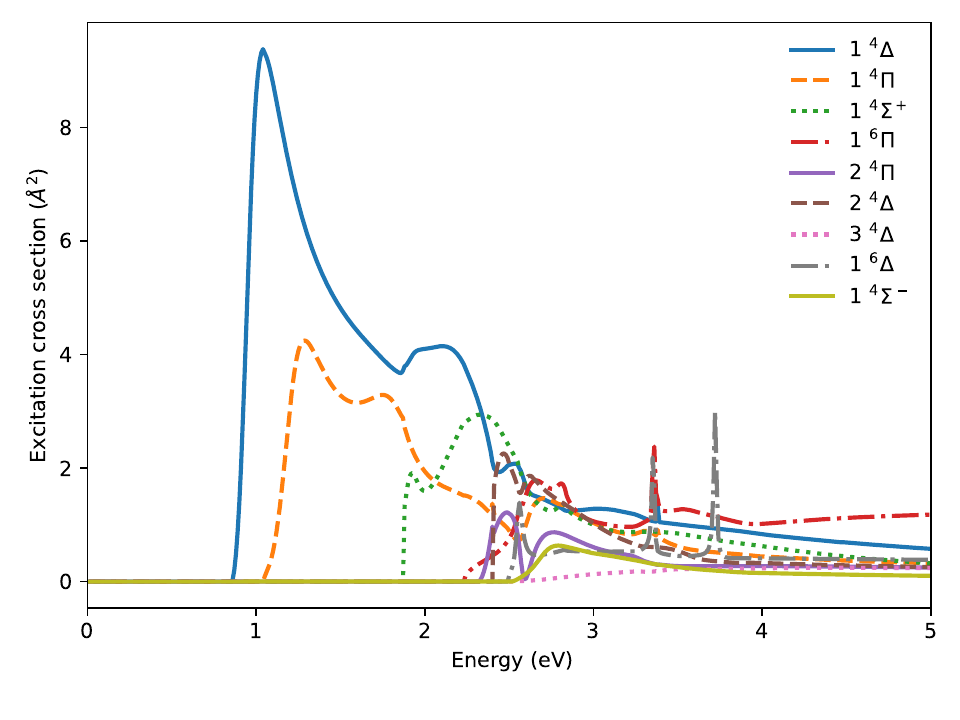}
    \caption{Electronic excitation cross sections for the lowest sextet and quartet excited states of WH.}
    \label{whex}
\end{figure}

Figure \ref{whex} shows the calculated excitation cross sections for the lowest few excited states of WH. The Born correction is applied to each cross section to account for high angular momenta. Given the high density of target states present in WH, which can increase the required computational resource significantly, the calculation presented here retains all states with vertical excitation energies below 4 eV. Higher lying states are likely to have negligible excitation cross sections over the energy range below 5 eV that is explored here.

\begin{figure}
    \centering
    \includegraphics[width=0.5\linewidth]{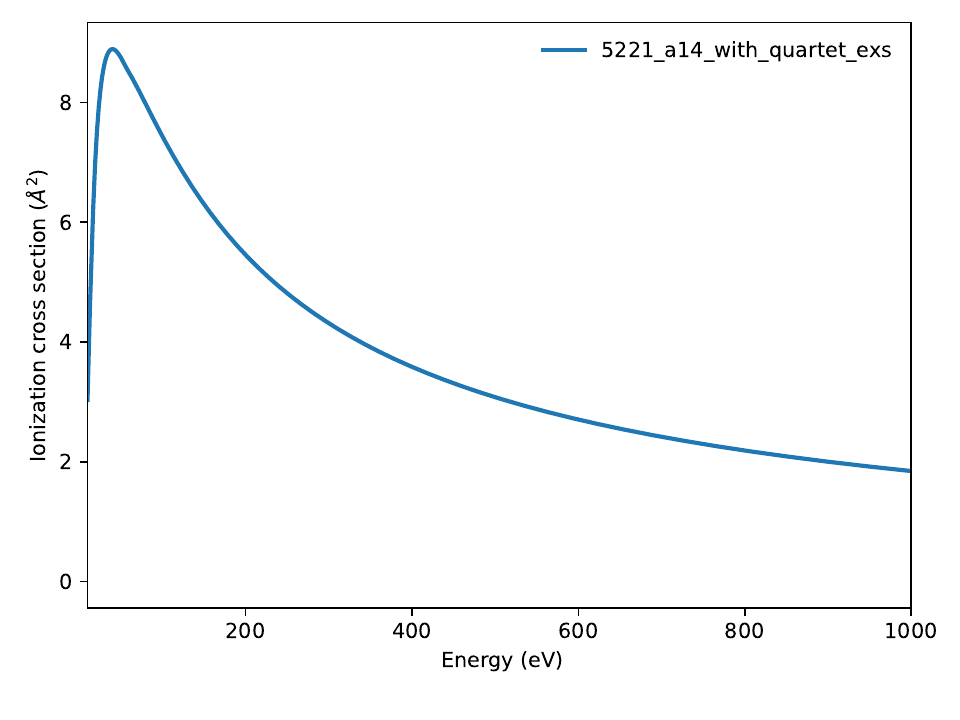}
    \caption{Total ionization cross section for WH.}
    \label{whi}
\end{figure}

Figure \ref{whi} shows the total ionization cross section for WH, calculated using the Binary-Encounter-Bethe (BEB) method \cite{kim1994}. The ionization energy given by Koopman's theorem is 6 eV.

\subsection{Photoionization of CH$_3$I}\label{sec:ch3i}

To complement the scattering calculations above, we include here as an illustration of the new functionality a photoionization calculation for the CH$_3$I molecule using the simple static exchange (SE) model.

The all-electron calculation used R-matrix radius of 30 a.u. and 45 B-splines of order 6 up to $l_{max} = 6$ and the $6-311G^{**}$ GTO basis set~\cite{glukhovtsev1995extension}. The hybrid 2-electron integrals were calculated using a combination of a semi-analytic approach for the $(BG|GG)$ class and the Legendre expansion with maximum angular momentum 20 for the mixed exchange integrals. The length of the elementary interval for application of the 21-point radial Gauss-Legendre quadrature was $\Delta r = 0.1 a.u.$

The calculations with ECPs replaced the innermost 28 electrons with the pseudo-potential optimized by Peterson et al~\cite{peterson2003systematically} alongside the cc-pVTZ-PP atomic basis set. The rest of the computational setup was kept the same as in the all-electron case.

\begin{figure*}[h!]
    \centering
    \begin{subfigure}[t]{0.5\textwidth}
        \centering
        \includegraphics[width=1\textwidth]{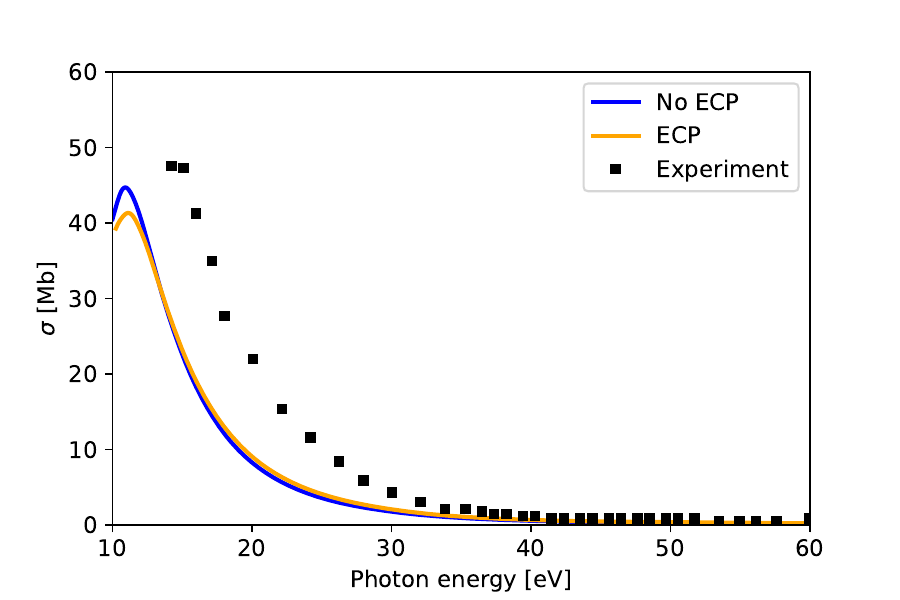}
    \end{subfigure}%
    ~ 
    \begin{subfigure}[t]{0.5\textwidth}
        \centering
        \includegraphics[width=1\textwidth]{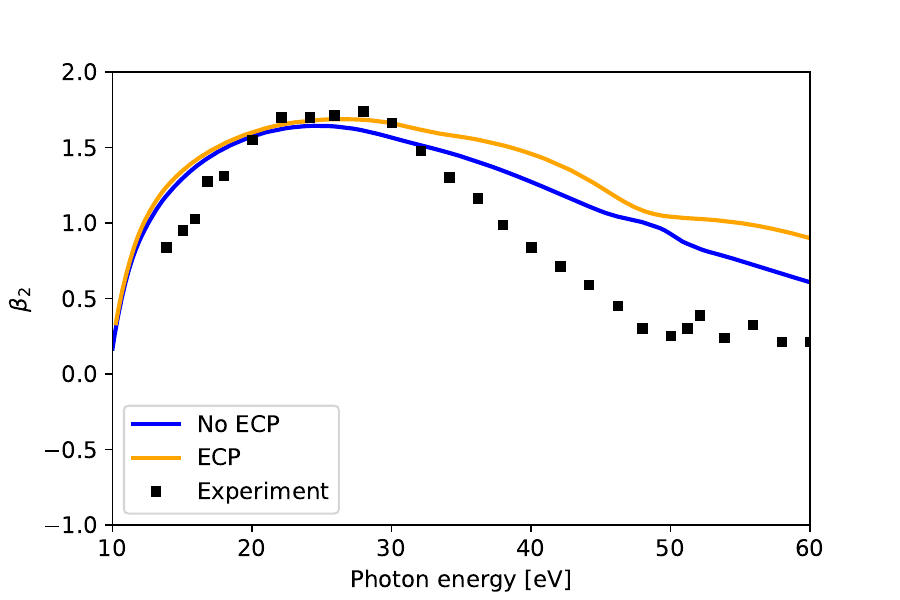}
    \end{subfigure}
    \caption{Photoionization of CH$_3$I into the ground state of the ion. Left: Photoionization cross section. Right: Photoelectron angular distributions. Calculations with ECPs are compared to all-electron calculations and experimental measurements \cite{holland2006study}.}\label{fig:photo}
\end{figure*}

Figure~\ref{fig:photo} shows the photoionization cross sections and the angular distributions ($\beta$-parameters) for the ionic ground state ($X^2 E$) from the two models in comparison with the experimental measurements of Holland et al~\cite{holland2006study}. The results with and without ECPs agree in general with each other except for the angular distributions at higher energies where the all-electron data appear to be slightly closer to the experiment. We see only a qualitative agreement of the theory results with the experimental data which is to be expected given the 1-electron nature of the SE model. The case of CH$_3$I is is complicated due to the Cooper minimum around $45$~eV and, at higher energies, the multi-channel effects of coupling to the 4d shell of iodine~\cite{holland2006study} and the spin-orbit splitting of the ionic ground state. Accurate description of those effects requires sophisticated modeling of polarization and electron correlation, together with inclusion of relativistic effects for the active electrons, which goes beyond the scope of this work.

\section{Conclusions}\label{sec:conclusion}

We have implemented Effective Core Potentials into the molecular scattering and photoionization code UKRmol+ and into the QEC interface. The implementation has been thoroughly tested for a range of molecules relevant for plasma modeling and demonstrated the ability of the codes to perform accurate calculations for targets containing heavy atoms. The new functionality is an important extension of the code's applicability since reliable all-electron bases for heavy atoms are typically not available, thus making all-electron calculations for those elements inaccurate and missing the important (scalar) relativistic effects.

Our implementation allows the use of ECPs with both GTOs and BTOs for the continuum description, the latter with a restriction to the semi-local ECPs. However, the semi-local ECP is often the only and dominant component of the ECP. Using ECPs scattering calculations for heavy atoms become often computationally comparable to calculations for light elements. Additionally, implementation of ECP integrals over BTOs allows us to perform scattering and photoionization calculations for extended energy ranges, up to 100 eV, as often required in various applications.

Future work will focus on inclusion of the spin-orbit ECPs, optimization of the momentum-space integrals over BTOs and on completion of the implementation of the local-type ECP integrals involving BTOs.

Implementation of ECPs also opens the route to representation of molecules embedded in extended environments with the effective electrostatic interaction represented by potentials of Gaussian type.

The approach based on ECPs naturally does not account for relativistic effects in the valence shell. Instead, those effects must be explicitly included using appropriate terms in the Hamiltonian~\cite{liu2020}. A promising route is the inclusion of the dominant relativistic effects in the non-relativistic Hamiltonian using perturbation theory, which can be conveniently combined with the R-matrix approach~\cite{benda2021}.

\section*{Data Availability}
All data that support the findings of this study are included
within the article. UKRMol+ is  open-source software which can be downloaded from zenodo.

\section*{Acknowledgement}

Development of Quantemol Electron Collisions (QEC) was supported by STFC grant ST/R005133/1. ZM, JB and MC gratefully acknowledge the support of the Czech Science Foundation (Grant no. 25-18015K) and of Quantemol Ltd. This work was supported by the Ministry of Education, Youth and Sports of the Czech Republic through the e-INFRA CZ (ID:90254).

\section*{Conflict of interest}
Anna Nelson and Jonathan Tennyson are Directors of Quantemol Ltd. Sebastian Mohr and Gregory S. J. Armstrong are Quantemol employees.

\section*{Author Information}

\subsection*{Corresponding Author}
 \textbf{Jonathan Tennyson} - Department of Physics and Astronomy,
University College London, WC1E 6BT London, U.K.;
https://orcid.org/0000-0002-4994-5238; Email: j.tennyson
ucl.ac.uk

\subsection*{Authors}
%
\textbf{Zden\v{e}k Ma\v{s}\'{i}n} 
https://orcid.org/0000-0001-9382-4655
\\
\textbf{Jakub Benda} 
https://orcid.org/0000-0003-0965-2040
\\
\textbf{Martin Crh\'{a}n} 
https://orcid.org/0000-0001-9414-0271
\\
\textbf{Gregory Armstrong} 
https://orcid.org/0000-0001-5949-2626
\\
\textbf{Anna Nelson} 
https://orcid.org/0009-0001-7741-5488









\providecommand{\latin}[1]{#1}
\makeatletter
\providecommand{\doi}
  {\begingroup\let\do\@makeother\dospecials
  \catcode`\{=1 \catcode`\}=2 \doi@aux}
\providecommand{\doi@aux}[1]{\endgroup\texttt{#1}}
\makeatother
\providecommand*\mcitethebibliography{\thebibliography}
\csname @ifundefined\endcsname{endmcitethebibliography}
  {\let\endmcitethebibliography\endthebibliography}{}

\end{document}